# Routing with Face Traversal and Auctions Algorithms for Task Allocation in WSRN

Jelena Stanulovic[1,2], Nathalie Mitton[3] and Ivan Mezei[1]

[1] University of Novi Sad, FTN, Trg Dositeja Obradovića 6, 21000 Novi Sad, Serbia; imezei@uns.ac.rs, jstanulovic@gmail.com
[2] Maxlinear Austria Gmbh, Europastrasse 8, 9500 Villach, Austria ; jstanulovic@maxlinear.com
[3] Inria, France, nathalie.mitton@inria.fr

**Abstract:** Four new algorithms (RFTA1, RFTA2, GFGF2A, and RFTA2GE) handling the event in wireless sensor and robot networks based on the Greedy-Face-Greedy (GFG) routing extended with auctions are proposed in this paper. In this paper we assume that all robots are mobile and after the event is found (reported by sensors), the goal is to allocate the task to the most suitable robot to act upon the event, using either distance or the robots' remaining energy as metrics. The proposed algorithms consist of two phases. First phase of algorithms is based on face routing and we introduced the parameter called search radius (SR) at the end of this first phase. Routing is considered successful if the found robot is inside SR. After that the second phase, based on auctions, is initiated by the robot found in SR trying to find a more suitable one. In the simulations, network lifetime and communication costs are measured and used for comparison. We compare our algorithms with two similar algorithms from the literature (k-SAAP and BFS) used for the task assignment. RFTA2 and RFTA2GE feature up to 7 times longer network lifetime with significant communication overhead reduction compared to k-SAAP and BFS. Among our algorithms, RFTA2GE features the best robot energy utilization.

## 1. Introduction

Wireless Sensor and Robot Networks (WSRN) consist of the combination of two types of wireless networks that cooperate: Sensor Networks and Robot Networks (sometimes called mobile actuators). A sensor network consists of static nodes with sensing capabilities, while a robot network consists of mobile nodes with actuating and enhanced capabilities. A WSRN system, based on these *cooperative* networks [1], can monitor the area for events using sensors, make a *localized* decision on which robot should move to the event location, and act upon a detected event(s). A cooperative WSRN network is illustrated in Figure 1 (blue dots are sensors, red dots are robots, blue lines illustrate the communication lines between the sensors, and red lines, accordingly communication lines between the robots). Let us assume an event E occurred near sensor S4, which sensed it and forwarded the information (including the event's position) to the event handler via sensors S3, S2, and S1. The event handler is an entity that collects the data from sensors and sends the order to robots. The event handler passes the information about the event containing the event location to a certain robot called the collecting robot (in this scenario R1). It relays the message through the robot network towards the robots in the event vicinity where local robots need to decide which one is the best to act.

The event handler forwards the information about the event containing the event location to a robot (in this scenario R1). To avoid time and communication costly service location discovery scenarios, the event handler chooses the robot to which it randomly sends the information by attaching its ID to the message with the event location. The chosen robot relays the message through the robot network towards the robots in the event vicinity where local robots need to decide which one is the best to act. In such a localized way, there is no need for both centralized knowledge and any robot location service discovery methods.

In some scenarios, the monitored event requires only additional information about the event without any action (e.g., video surveillance of the event location). In other applications, such as smart indoor monitoring of buildings (e.g., temperature, smoke density, air humidity, vibrations, etc.), upon the detection of the problems (e.g., smoke is detected in some area of the building), a robot is expected to react immediately and resolve the problem. Robots are particularly useful in scenarios where human participation is dangerous or not possible (e.g., rescue activities after natural disasters [2]).

One of the main problems in the wireless sensor and robot networks is how and to which robot(s) to allocate the task. In this paper, it is assumed that the sensors received the information about an event that happened somewhere in the network and the best robot is supposed to react. Accordingly, the task assignment problem (also known as the task allocation problem) is to find the most suitable robot to react to the event. This problem is in robotic literature well known as a Multirobot task allocation problem and in [3], [4], and [5] are reviewed and presented the existing solutions and their variations in WSRN.

In this paper, we tackle the task allocation problem. We proposed four new solutions based on greedy face traversal routing extended with auctions. These new routing algorithms rely on the extension of the GFG routing [6] proposed in our previous work [7]. The motivation for the new algorithms came from a previous work [7] that showed the need to improve the GFGF algorithms in terms of higher efficiency, lower communication costs, and longer network lifetime. The basic scenario on which focus in this paper, is the following. The sensors identify the event location and the location information is sent to one of the robots (robot R1 in Figure 1). The event is somewhere in the area monitored by the robots, but these latter need to move to the exact location to perform the task. We use the GFGF2 algorithm to find a robot near the event location in the search radius SR, as in [7], and extend it with auctions to find whether there is a better robot that can perform the task compared to the one found within search radius SR around the event. Besides, while in [7] the focus is on finding the closest robot to react to the task, here we aim to find the most suitable robot to react based on the metrics that take account of the remaining energy, the distance to the event, and the speed of robot movement.

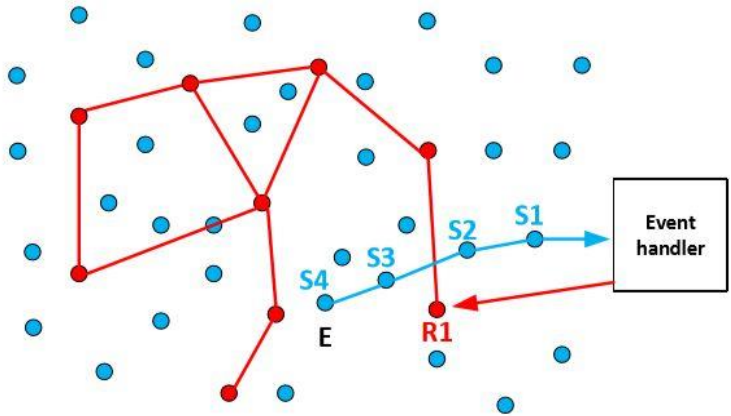

**Figure** 1: Cooperative WSRN

Auctions used in this work are based on k-SAP auction protocol [8]. The first proposed algorithm uses distance metrics and is called RFTA1. The other proposed algorithms, RFTA2, GFGF2A, and RFTA2GE, use the remaining robot energy as the metrics. Here, we measure the network's lifetime, energy balancing among robots, and robot utilization.

The main contributions of this paper are the following:

- We improve the existing greedy face greedy based routing solution and apply it to the robot task allocation problem. It resulted in a longer network lifetime and better energy balancing using a combination of face traversal routing and auctions based on different metrics evaluated for different scenarios and network topologies.
- We introduce the RFTA2 algorithm, which shows a network lifetime up to 7 times longer compared to k-SAAP [28] and BFS [29] algorithms with significantly less communication costs. It also shows a network lifetime up to 5 times longer compared to the GFGF2A algorithm (for little additional communication costs).
- We introduce RFTA2GE, which features the best robot energy utilization and energy balancing among all robots for additional communication costs. Since the communication costs are usually by order of magnitude lower than robot movement costs, this is a highly beneficial contribution of RFTA2GE. It features a network lifetime up to 7 times longer than k-SAAP [28] and BFS [29] for 3 times less communication costs.
- Within the algorithms RFTA2 and RFTA2GE, we introduce the parameter SR (Search Radius). It is the radius of the circle which defines the space around the event where the search for robot is performed. We also determine its optimal value and prove it mathematically.

The rest of the paper is organized as follows. The overview and explanation of the related work protocols are given in Section 2. In the Section 3 the system model is presented, and the newly proposed algorithms are presented in the following section. In Section 5 are presented simulation results and a discussion of the obtained results. Section 6 concludes the paper and provides future work proposals.

## 2. Related work

Here we present related work on the geographic routing protocols that are the basis for our work presented in this paper. It is face routing [9], Greedy routing [10], Greedy-Face-Greedy (GFG) routing [6], Greedy-Face-Greedy-Find (GFGF) [7] and auctions [8]. Besides, we review the literature related to the task allocation problem.

### 2.1 Task allocation

In the multi-robot systems literature, task allocation is a well-studied topic (e.g., the survey [11]). The task allocation strategies can be classified according to their applications [12]. Optimization- and auction-based techniques are described as the most important task allocation strategies ([13], [4] and [5]).

Task allocation aims to match appropriate robots and certain jobs that are required by the system most efficiently [14]. The auction-based task allocation approach is one of the possible solutions whereby using auctions, robots are bidding on a task. The bids are based on certain metrics, and the cheapest one is assigned to the task [15]. Another approach, presented in [16], is the greedy approach and is implemented using the closest robot. In our paper, we are using a combination of these two approaches.

Additionally, another feature to consider is the connectivity of the network with the task allocation problem, which is studied in [17]. The authors in [18] investigate the task assignment problem when robots are constrained by a limited communication range. They propose both centralized as well as decentralized solutions. In contrast to many task allocation approaches that assign one task at a time per robot, in [19], single robots are performing multiple tasks simultaneously with different priorities.

The geographical routing in the WSRN has become an important research topic. In location-based routing, since the decisions are made based only on location information, there is no need for complex computation. As location-based protocols do not use the information about full topology but only location information, they are very efficient in terms of routing. A survey and taxonomy of location-based routing for WSN are presented in [20]. They categorized the recent research works into several directions.

In order to improve efficiency and lifetime of the network auctions are used in [28]. They use the localized auction aggregation protocol (k-SAAP). It is shown that the proposed algorithm in some scenarios guaranties 100% efficiency in finding the closest actuator.

RODAA was presented in [29] and the goal of the paper was to optimize robot resources and task completion time. It is based on BFS algorithm [32]. The BFS tree is used to overcome the weakness of the auction algorithm, in which only the single-hop neighbors of the auctioneer can participate in the auction process if the communication range of the robot is limited. Since the tasks must be allocated in a dynamic environment where the robots are mobile and the tasks occur anywhere at any time, the BFS tree-based approach guarantee conflict-free task allocation.

In [30] is proposed an adaptive auction protocol AAP for task assignment in multi-hop wireless actuator networks, and the scenario that is considered is where all actuators are static and each of them can obtain the target information from sensors. Unlike existing methods that neglect the adaptive auction area required by dynamic networks, the proposed method uses an adaptive factor that is deduced based on the relation between network characteristics and protocol performance. The robots is considered to be static so this algorithm is not suitable for comparison with our algorithms.

*2.2 Face routing*

This type of routing assumes that the network is divided into faces. Face routing protocol [9] forwards the message along with the adjacent faces. Those faces are all intersecting the line connecting source and destination *sd*. When the message reaches the edge, which is intersecting the *sd* line, the face is changed. Only intersections that are closer to the destination, compared to the last intersection point, are considered. There are several variations of the algorithm called *before crossing*, *after crossing*, as well as *the left-hand* and *right-hand* rule. The left-hand rule means that the message is traversed along the edge that is placed counterclockwise from the previously examined edge. The right-hand rule assumes that the message travels along the edge, lying in a clockwise direction.

The before/after crossing variants are used for the next face selection. Using before-crossing variant, the message is forwarded along the edge which is not intersected by the line connecting source and destination. In the other variant, the other edge is selected, i.e., the one intersected by the line *sd*. The third possibility is called *the best angle variant* minimizing the angle between the possible edges to choose.

*2.3 Greedy routing*

Greedy routing [10] is one of the simplest protocols for geographical message forwarding. Using this protocol, the message is continually forwarded closer to the destination. The message routing is started from a source node (e.g., robot) towards the destination node at a specific location. To decide where to forward the message, the sender checks which of his neighbors are closer to the destination than itself. Accordingly, the next node is the new source performing the same action to forward the message further. This operation is continued until the destination is reached or until none of the neighbors are closer to the destination than the current source. In such case, it is not possible to forward the packet; routing fails, and the packet is dropped. This is the main drawback of the greedy routing. There are several recovery techniques and protocols (e.g., GFG).

*2.4 Greedy face greedy (GFG) routing*

Greedy face greedy routing (GFG) [6] is based on the combination of greedy and face routing. In the greedy phase, the packet is forwarded along with the nodes in such a way that the next node is the closest of all potential neighboring nodes to the destination compared with the current node. When there are no closer neighbors, the face phase of the algorithm is started, and it is performed until the closer node is found. The main idea is that the message is forwarded to the next node on the face until the intersection between the line connecting the source and destination node and the line connecting the current and next node is found. If such an intersection is encountered, the face is changed and the message is forwarded to the node in the next face. The GFG algorithm is switching between greedy and face phase as many times as needed until the delivery is performed since this algorithm is routing with guaranteed delivery.

2.5 *Greedy face greedy find (GFGF) algorithms*

Greedy face greedy find (GFGF) algorithms are based on GFG when it is applied to the scenario where destination D (event location) is not part of the network [7]. There are two algorithms designated as GFGF1 and GFGF2. The main issue is the definition of the stop criteria since the destination is not part of the network and cannot be reached with the standard GFG, and accordingly, the routing would never finish, and the event would never be found. GFGF1 routing stops when the same edge of a face has already been visited. In that way, with this algorithm, the event is surrounded by the nodes in the face around it (e.g., event D1 in Figure 2). Also, it is shown that the node closest to D could always be found. Furthermore, in the case where the event is somewhere around the outer face of the network (e.g., event D2 in Figure 2), GFGF1 routing has high communication costs since the face around the event is the outer face. GFGF2 algorithm is used to cope with this problem when the event is outside of the network. The difference to the GFGF1 is the stop criteria.

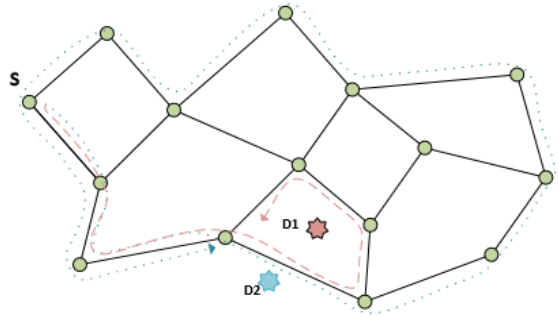

**Figure 2** Two scenarios: D1 in and D2 outside of the network area

Both possible scenarios for the GFGF2 are illustrated in Figure 3. If the robot could not be found within the search radius SR, the routing would continue all around the network (following the so-called outer face). When it comes back to the node where it started, the routing stops, even though it is not inside the search radius SR around the event. In the search scenario with search radius 2SR, there is a robot inside, and the stop criterion is satisfied with less communication overhead. The closest robot is found in 55-90% of routings, compared to GFGF1 where it is almost 100% [7].

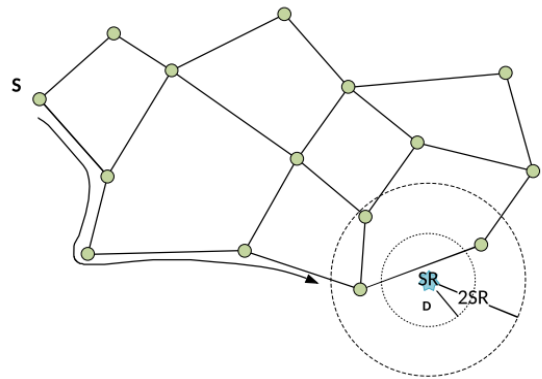

**Figure 3** Illustration of the GFGF2 algorithm

2.6 *Auctions*

The market-based approach and auctions are often used in robotics to do task allocation [11]. However, the complete network communication graph (i.e., each robot can communicate with each other) is often assumed, which is a hard constraint. Instead of flooding the whole network, localized multi-hop auctions called k-SAP are introduced in [8]. The auctions are organized by a robot called auctioneer, which sends a message to all of his first neighbors. The ones which, according to the used metric, are able to do the task are bidding back. The bids are searched only among the robots which are located within k-hops from the auctioneer. Then the auctioneer decides which robot is the best to run the task. The example of a metric based on multi-objective goals is given in [22].

### 3. System model

We assume that the network consists of two types of nodes: static sensors and mobile robots. Both sensors and robot nodes are assumed to be aware of their position. The sensors obtain information about the location of the event and a robot is supposed to be assigned to react upon that event. All the robots in the network have the same transmission radius *r*, which means that each robot can communicate only with its first robot neighbors, located inside the radius *r*. Accordingly, the robot network is connected all the time but does not form a complete communication graph (i.e. not all robots can directly communicate with each other) since robot communication is modeled as a unit disc graph (UDG). Communications among robots are done in a multi-hop manner. The communication between every two robots is possible by sending the communication messages over other robots which only relay messages towards the destination. We assume that robots and sensors cannot communicate.

We assume that all of the robots know their position (e.g., by having localization hardware on itself). The collecting robot is the one that gets the information about the event location and informs the other robots about the event. The typical solution of the task allocation problem is centralized, assuming a complete communication graph, which can be optimal but needs complex computations (e.g., by using linear integer programming [21]). In this paper, we propose several localized solutions based on a multi-hop communication model. In such a model, the collecting robot that gets the information about the event (from outside of the robot network by sensors), is considered the source and it starts the routing towards the other robots (e.g. those closer to the event). The message containing the location of the event is then forwarded through the robot part of the network. We consider random and random topologies with hole. To perform the message routing and eventually allocate the robot for the task on the event destination, we focus on the algorithms based on greedy face traversal routing combined with auctions.

To reduce the complexity of routing algorithms, only some of the available communication links are considered. To do so, we need to apply a graph reduction tool that maintains connectivity. Compared to the others, the unit disc graph (UDG) has no restrictions other than a fixed transmission radius but on the other hand has a high interference due to the high number of the connections. The strongest restriction that does not affect and even preserves the connectivity is based on a spanning tree. However, it features a low diversity of routing paths and therefore has less possibility for energy balancing and routing optimizations. Another solution that has many favorable features is the Gabriel graph (GG). The latter has significantly lower number of the connections, which reduces communication overhead, and collisions compared to UDG. In addition, it induces a planar graph, contains the Euclidian minimum spanning tree as a subgraph, and has optimal energy spanner [23]. On the other hand, the relative neighborhood graph (RNG) removes more edges and has lower diversity in paths compared to GG. The degree of the graphs (average number of the neighbors) is in the following relations:

$$d(UDG) > d(GG) > d(RNG)$$

Accordingly, there are fewer routing paths for the RNG compared to GG. On the other hand, GG features a lower degree than UDG, accordingly having less communication links than UDG. For those reasons, we use the Gabriel graph of random UDG as a middle solution in terms of the number of connections compared to UDG and RNG to model the robot network in this paper.

**Definition**: Task allocation problem

For a given set S of m sensors, S = {$S_1$, $S_2$,… $S_m$}, a given set R of n robots, R = {$R_1$, $R_2$, …$R_n$}, with known locations and initial energies of robots, given a set of e successive events E = {$E_1$, $E_2$, …$E_e$} detected by sensors, with known locations, the problem to be solved is to find the best robot to react upon the event and to be assigned the task for all consecutive events.

Here is the list of assumptions:

- All the robots in the network have the same transmission radius *r*, and each robot can communicate only with its first neighbors, located inside of the radius *r*.
- All the robots know their position and the position of the current event is known. Other events are not known *a priori*.
- All the robots have a certain amount of energy at the beginning of the routing process called *initial energy*. In the beginning, we assume it is at 100%. While performing the task, the robot energy is drained proportionally to the distance traveled.
- The search radius SR is a circle around the event where the robot is searched; if any robot is identified, the routing stops. Since the robots can determine both their position and the position of the event, they can decide locally whether they are in the circle or not.
- During auctions, robots calculate the energy needed to do the task. If it does not have enough energy to do the task, the robot does not bid back.
- To evaluate the algorithms, we focus on communication costs and the network lifetime.
- Communication costs are based on the number of messages needed to route the message from the source to the destination during one round of simulations (i.e. for one event).
- The network is considered alive until a task cannot be allocated because none of the neighboring robots have enough energy to perform it, and no one bided.

**4. RFTA2 algorithms**

We propose a family of four new algorithms, designated as RFTA (Routing with Face Traversal and Auctions) to further improve already suggested GFGF routing algorithms. Basic idea is to improve their efficiency using auctions. Four new algorithms are described in the following subsections. All of them are two-phase algorithms. First, Greedy-Face-Greedy-Find (GFGF) routing is active until one of the stop criteria is reached. It is followed by the second phase – the phase with auctions. In the auctions phase, the last robot who got the message with the GFGF algorithm is the one organizing the auctions (i.e., this robot is the auctioneer). Auctions are always performed in the way that the auctioneer sends one message to all of its closest neighbors (i.e., 1-hop neighbors). Neighbors bid according to the criteria based on either the distance to the event (RFTA1) or the remaining energy the robot would have left after doing the task (RFTA2 and RFTA2GE).

*4.1 RFTA1 algorithm*

The RFTA1 algorithm consists of two phases. In the first phase, the routing is done using the GFGF2 algorithm. The auctioneer sends the message about the event to all of its neighbors, as well as its bid, which is assumed to be the best at the beginning of the auctions. To be communicationally efficient (i.e., low number of messages), only the robots that are closer to the event destination D (task) than the auctioneer send their bids. The auctioneer collects all the offers and decides which robot is the best to run the task, based on the distance to the event location D.

Pseudo code for RFTA1 algorithm:
1. **Do** GFGF2 **until** a robot within the search radius SR around the event destination D is found **or until** the next node has already been examined (in that case, take the last examined robot as the winner)

2. **Assign** that robot to be the auctioneer
3. Auctioneer: Start auctions
   - The auctioneer sends the message to all its first neighbors, calling for the closest to the event destination D
   - Robots bid back if they are closer to the event
   - The auctioneer decides which is the best to respond and allocate the task

*4.2 RFTA2 algorithm*

In this algorithm, the remaining energy of the robot's battery is used as the metric. The first approximation for the energy that is consumed for the robot movement is based on the 'linear rule' [24], where the energy consumption is calculated as a linear function of the distance. In this paper, each of the neighbors' bid is based on the following calculation which takes account of the remaining energy they have, their distance to the event location D, and their speed, using the Equation (1) taken from [25],

$$EnergyLoss = 6.25 \cdot vRobot \cdot d + 9.79 \cdot d + 3.66 \cdot \frac{d}{vRobot} \qquad (1)$$

A robot calculates the remaining energy in case it is considered to be allocated for the task (distance d is sent as auction parameter by the auctioneer). The robot bids only if it estimates to have enough energy to run the task. The auctioneer decides which robot is the most suitable for the task.

There are four possible cases to make the decision and allocate the task to one of the robots. First, if more than one of the robots bid for the task, the auctioneer chooses the one that would have the most remaining energy left after finishing the task. In the second case, if only one robot bids for the task, it is assigned the task. The third case assumes that if none of the robots bid due to low remaining energy, they are unable to perform the task. In that case, the auctioneer assigns the task to itself (provided that it has enough energy). The last, fourth, case would be when no robot bids for the task and the auctioneer robot also does not have enough energy to perform it. It is recognized as the stop criterion for the RFTA2 algorithm and, consequently, network lifetime is considered exhausted.

4.2.1 Update of the network

Once a robot is assigned a task, it moves to the task coordinates (i.e. event destination D), and the network is updated. The update means that the neighbor robots are recalculated to fulfill the communication radius criteria and Gabriel's graph.

In this way, the routing can be done in rounds as long as there is a robot to react, or all robots have enough remaining energy. With this criterion, the algorithm stops if in the current round there is no robot to react.

Figure 4 shows the part of the network wherewith first part of the RFTA2 algorithm, a node inside the radius SR, is reached. This node is assigned to be the auctioneer A. It sends messages to its first neighbors (R1, R4 and R2) illustrated with blue arrows. Node R2 is the one bidding to perform the task; its message is sent back to the auctioneer and illustrated with a green arrow

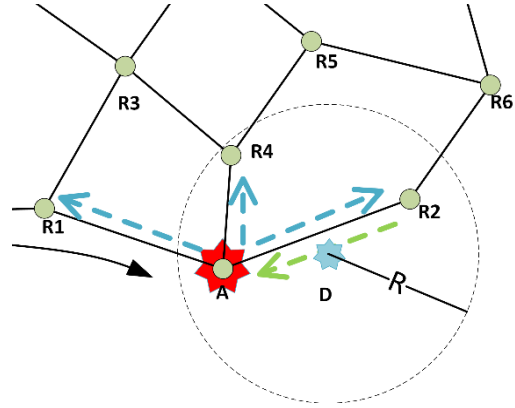

**Figure 4** Illustration of the RFTA2 routing

---

Pseudo code for RFTA2 algorithm:

1. **Do** GFGF2 **until** a robot within the search radius SR around the event destination D is found or **until** the next node has already been examined (in that case, assign the last examined robot as the winner)
2. **Assign** that robot to be the auctioneer
3. Start auctions
   - The auctioneer sends the message to all of its first neighbors asking for bids to perform the task at the event destination D
   - Robots calculate their remaining energy and bid accordingly
4. The auctioneer decides to which robot to assign the task based on the bids
   - **If** #bids > 1
     - The auctioneer assigns the task to the one with the most remaining energy after completing the task
   - **Else If** #bids==1
     - The single bidding robot is allocated the task
   - **Else**
     - **If** the auctioneer has enough energy, it takes the task, **otherwise**, stops
5. The robot who has been assigned the task moves to the coordinates of the task and the network is updated
6. Start new round, **goto** 1 and **repeat** steps 1-5, **otherwise** stop when there is no robot to react due to the low remaining energy.

.

*4.3 GFGF2A algorithm*

GFGF2A is the extension of the GFGF2 algorithm, developed to fairly compare the new RFTA algorithms and the GFGF algorithms. The extension is based on the assumption that all the robots start the routing sequence with the same energy level (assumed to be at 100%). When the algorithm stops because one robot in the radius SR around the event destination D is found, that robot moves to the coordinates of D in order to perform the task. The energy that the robot consumes to perform the task is calculated based on the Equation (1).

---

Pseudo code for GFGF2A algorithm:

1. **Do** greedy **until** delivery or failure
2. **If** failure **then**
   - Search the next node on the face, based on the right-hand rule
   - **If** (node is within the radius SR around the event destination D)
     - finish ends
3. **Repeat** steps 1 and 2 **until** the node within the search radius SR is found or **until** the next node has already been examined

4. The last robot reacts, and its energy consumed to perform the task on destination D is calculated
5. The robot which is assigned the task moves to the coordinates of D and the network and its neighbors are updated
6. Start new round, **repeat** steps 1-5
7. **Stop** when there is no robot to react due to the low remaining energy

*4.4 RFTA2GE algorithm*

To improve RFTA2, we propose the greedy extension of this algorithm designated as the RFTA2GE. It is based on the method proposed in [26]. After the reacting robot is chosen, its first neighbors are examined to determine whether some of them have energy left after potentially running the task in the destination D. Auctions are started as in RFTA2 and the reacting robot chooses some of the neighboring ones to react. Its first neighbors are examined as in RFTA2 but also their first neighbors are informed about the event and they bid as well if they can offer to do that task on the coordinates of the destination D.

The idea is that this greedy extension improves the performance of the RFTA2 in terms of longer network lifetime, but at the cost of extra communication.

Pseudo code for RFTA2GE algorithm:

1. **Do** GFGF2 **until** a robot within the search radius SR around the event destination D is found or **until** the next node has already been examined (in that case, assign the last examined robot as the winner)
2. Assign that robot as the auctioneer
3. Start auctions
    - The auctioneer sends a message to all its neighbors to bid for the task in destination D
    - Robots check whether they have enough energy for that and bid back
    - Each of the neighbors sending the message to all their first neighbors and bid back in case they have enough energy to do the task
4. The auctioneer selects the robot to react based on the bids:
    - **If** #bids > 1
        - The one with the most remaining energy after doing the task is chosen.
    - **Else If** #bids==1
        - The single bidding robot is allocated the task
    - **Else**
        - **If** the auctioneer has enough energy takes the task, **otherwise**, stops; **goto** 7
5. The selected robot moves to the coordinates of the task and the network is updated
6. The new round is started, **repeat** the steps 1-5
7. The algorithm ends when the robot to react has not enough energy to run the task

Figure 5 shows part of the network where using the first part of RFTA2GE algorithm reaches a node inside the radius SR. It is assigned as the auctioneer A. It sends messages to its first neighbors R1, R4, and R2 (blue arrows). The greedy extension is depicted with orange arrows where messages are also sent to the second neighbors (in the part of the network on the Figure 5 visible are R3, R5, and R6). Green arrows show one node that bids back to perform the task at the location D. In this example, that is node R6 and it is sending a message to R2 which is further informing the auctioneer.

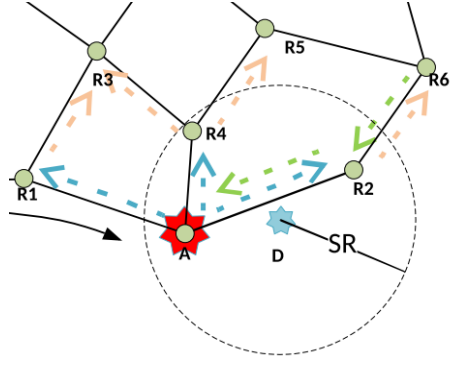

**Figure 5** Illustration of the RFTA2GE routing

*4.5. Optimal radius value for RFTA2 and RFTA2GE algorithms*

In RFTA2 and RFTA2GE, the search radius SR will be changed from SR to 4SR in SR steps (where SR = 0.1). It can be shown that the best results in terms of network lifetime are obtained for the value of 2SR. Here we present Lemma 1 and Lemma 2 to prove that.

*Lemma 1*
*The lower bound of the probability P(R) of finding at least one random point within a circle with the radius R (around one random point) and center is within the unit square is given by (2)*

$$P(R) \geq 1 - (1 - \pi * R^2/4)^{N-1} \qquad (2)$$

*where N is a sufficiently high number of random points within the unit square.*

**Proof of Lemma 1:**
Using geometrical probability, the worst case of finding one random point out of N random points within a circle with the radius R in the unit square is the case when the random point is in the unit square corner. If we express the probability $P_{out}$ of all other N – 1 points that are outside of this circle in the corner n $P(R) \geq 1 - P_{out}^{N-1}$, where $P_{out} = 1 - \pi*R^2/4$ is the probability that one point is outside of this circle.

*Lemma 2*
*For the set of N random points in unit square and the set of search radius {SR, 2SR, 3SR and 4SR; SR=0,1}, the best value (i.e., incurring the highest network lifetime) for the search radius used in RFTA2 and RFTA2GE algorithm is 2SR.*

**Proof of Lemma 2:**
For a sufficiently high number of points, e.g. N = 100, according to Equation (2), P(0,2) = 0,96 means that the probability of finding at least one robot is more than 95% and accordingly for 2SR case the best results are to be expected. To prove it, we show that the other 3 cases (SR, 3SR, and 4SR) feature worse results. According to Equation (2), P(SR=0,1) = 0.54 thus finding at least one robot within a circle with search radius SR=0,1 is not probable in almost half of the cases, and routing fails in those cases. Besides, as holds P(3SR=0,3) = P(4SR=0,4) = ~ 1, at least one robot is (almost) always found. However, due to the larger search radius, the probability that more than one robot is found in it is accordingly increased. Since most of the routings start from the outside of the circle around the event, it encounters the first robot within a circle positioned closer to the circumference than to the center of the circle (event location). This can be also confirmed by the fact that the probability of having a random point in the outer half of the circle area (probability is 1-1/4) is three times higher than within the central half of the circle area (probability is

π(SR/2)2/πSR2 = 1/4). Accordingly, the routing stop criterion is met earlier and the robot is at a larger distance to the event, accordingly spending more energy to come to the event location and thus reducing the network lifetime.

*4.6 Complexity analysis*

To analyze the complexity of our algorithm we are proposing two Lemmas.

*Lemma 3 (RFTA2 upper bound)*

*The number of routing steps for RFTA2 and RFTA2GE are upper bounded by O (n\*c) where n is the number of robots and c is the number of edges representing communication links.*

**Proof of Lemma 3:**

In [9] it is shown that GFG is upper bounded by O (n*c) provided that left-/right-hand rule is not changed while exploring faces, a message explores a face at most once and then the algorithm stops or message is advanced to another face, and all visited faces are mutually distinct. This assumption holds for our face traversal strategy. Auctions, as well as greedy extension, are in O(1) since it requires a constant number of rounds: 3 communication rounds (1 round for information, 1 round for biding, and 1 round for the task allocation) and 6 routing steps (2 rounds for information, 2 rounds for bidding, and 2 rounds for the task allocation), respectively. Accordingly RFTA2 and RFTA2GE are upper bounded by O(n*c).

*Lemma 4 (communication complexity upper bound)*

*RFTA2 communication complexity is upper bounded by $O(n^2+n)$.*

**Proof of Lemma 4:**

Let M - number of messages for the task allocation

F - number of internal faces of the graph

deg- average number of neighboring nodes, deg ~ $(n-1)*(\pi R^2/A)$ where A is   area being monitored

From [33] F = c – n + 1, and from planar graph theory it is well known that n-1 < c < 3n

M = (F + 1)*deg + 1 + 1 = (c-n+2)*deg + 2

O(M) = O(c-n+2)*O(n) = O(c*n-n*n+2n)=$O(n^2+n)$

Accordingly, communication complexity of M is upper bounded by $O(n^2+n)$

The comparison of ours and the algorithms from literature based on communication complexity, robot mobility, method and simulation environment are given in Table 1.

Table 1 Analysis of task allocation algorithms

| Algorithm | Complexity | Mobility | Method | Simulation env./source code |
|---|---|---|---|---|
| BFS [29] | O(diam · \|E\|) | yes | Breath-first search | robot simulator Webots /not available |
| k-SAAP [28] | N/A | yes | Localized Auctions with k neighbors | C programming lang./available |
| AAP [30] | N/A | no | Localized Auctions with adaptive number of neighbors | OMNet++ simulation tool /not available |
| MIA-TA [31] | $O(n^2)$ | yes | Auctions with maintaining network connectivity | Python on a PC with AMD 4.1 GHz CPU and 8GB RAM /not available |
| RFTA2 | $O(n^2+n)$ | yes | GFG with auctions | C programming |

**5. Results and Discussions**

The simulation scenarios are written using the C programming language. Network parameters used in simulations for all algorithms are as follows. Random and random with hole connected network topologies based on random Gabriel graph are used. In both topologies, the nodes are uniformly distributed in the field. The monitored area was a unit square, the node coordinates were between 0 and 1, the network consisted of 100 nodes (robots), and the communication radius $r$ is varied from 0.15 to 0.55 in steps of 0.05. According to [27], there is a sharp edge for $r < ((\ln n)/\pi n)^{1/2} = r_C \sim 0.12$ below which the network is not connected (which is also confirmed by our simulations). Hence, we set a lower bound for the range to 0.15. For values higher than 0.55, network becomes too dense, approaching a complete graph. Each round of simulations is repeated 100 times. Results are presented as averaged values of 100 simulations. In the graphs we show the simulation results together with their 95% confidence intervals.

For RFTA1 algorithm average communication costs and percentage of finding the closest robot to the task were measured and compared to GFGF2. The results are shown in the Figures 6-8. The results show that RFTA1 has higher communication costs in terms of the number of messages compared to GFGF2 with almost the same success rate of finding the closest robot for the task. It seems that introducing auctions does not bring any benefit for both topologies.

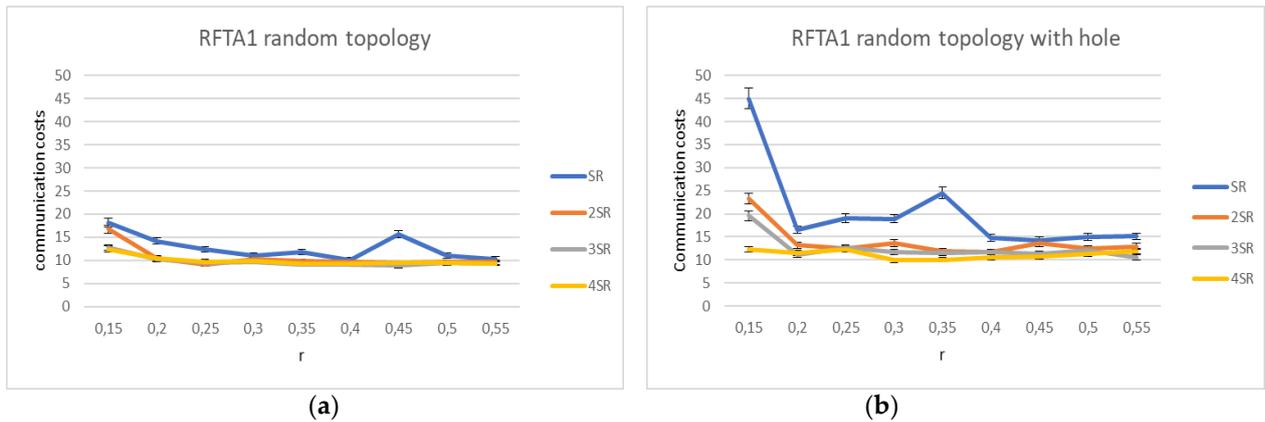

**Figure 6** Communication costs for the RFTA1 algorithm for (**a**) random topology (**b**) random topology with hole

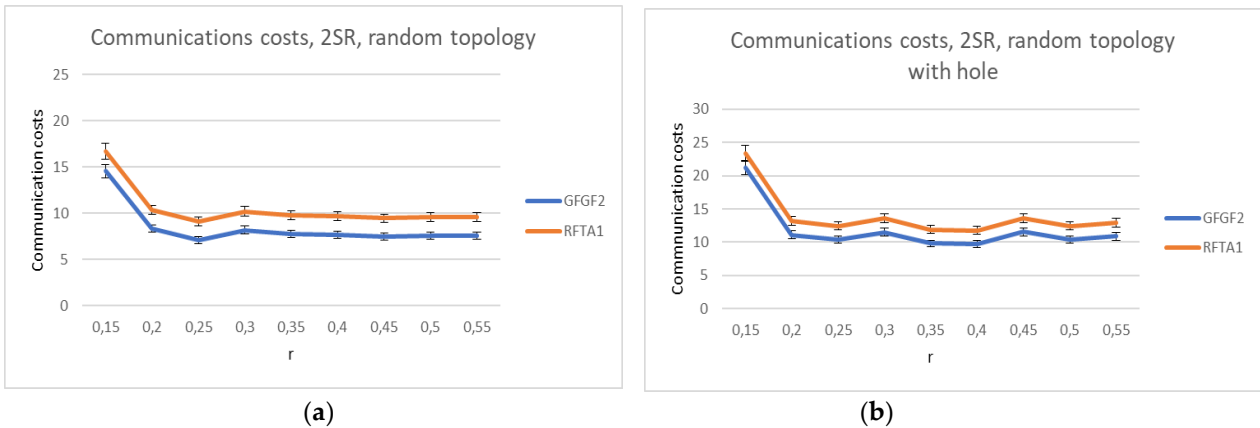

**Figure 7** Communication costs comparison for the RFTA1 and GFGF2 algorithms for (**a**) random topology (**b**) random topology with hole

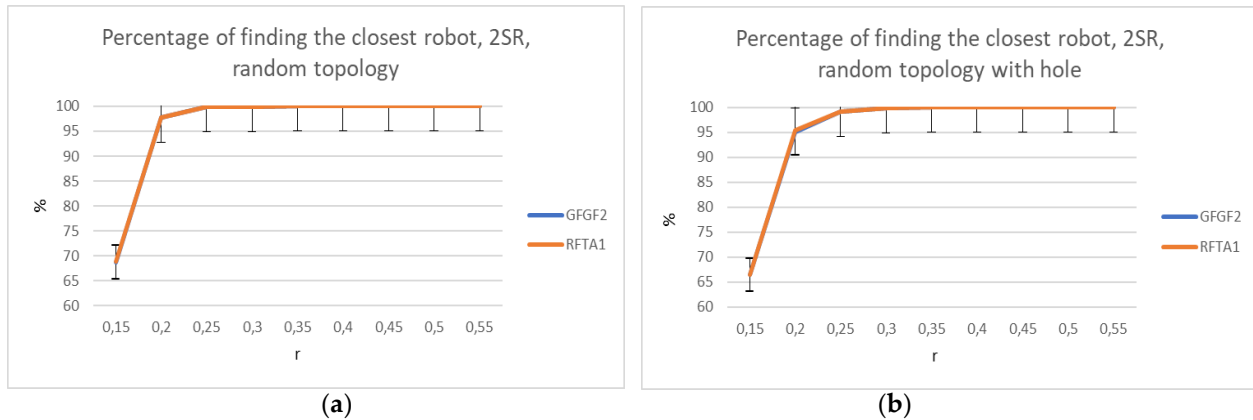

**Figure 8** Percentage of finding the closest robot, comparison for the RFTA1 and GFGF2 algorithms for (**a**) random topology (**b**) random topology with hole

For RFTA2 algorithm, the measured values were communication costs, in terms of the number of messages, the lifetime of the network, measured in rounds, and the remaining energies of the robots after the algorithm stop criterion is met. The stop criterion is met when none of the robots participating in the auctions have enough energy to do the task.

Simulations are executed in rounds. In each simulation round, one event occurs at random coordinates, which are not part of the network. The speed of the robots is assumed to be constant $v = 0.76\ m/s$, as in [25]. For distance calculation, the coordinates of the nodes were multiplied by 10 and assumed in meters.

The algorithms proposed in this paper are compared with our algorithms previously presented in [7] as well as with two algorithms from literature k-SAAP [27] and BFS [28]. For both algorithms, the bid metrics is adapted to be based on the energy and the distance proposed according to Eq. (1) from [25]. In that way all algorithms are using the same bidding metrics for fair comparison. Values of k in k-SAAP and hopmax in BFS are chosen based on the thorough simulations. For k-SAAP it is 7. For lower values, the network lifetime is very low. The reason is the distance from the auctioneer to the destination and that for small k there is almost never a suitable robot. If the k value is too big, the communication costs are much higher while network lifetime improvement is almost negligible. For the BFS algorithm, different values of hopmax were set to 7 and 10. Figure 9 show total communication costs used for the task assignment obtained in simulations averaged for 100 random networks.

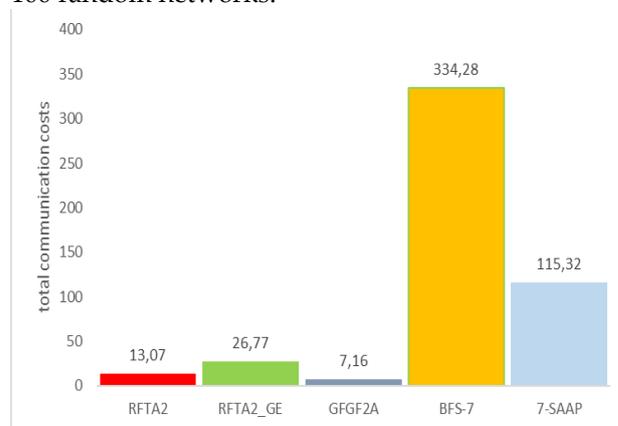

**Figure 9**: Total communication (for 2SR and r = 0.2)

From the simulation results presented in Figure 10 it can be seen that RFTA2 and RFTA2_GE feature a lifetime around 7 times higher than 7-SAAP and BFS-7. Communication costs for 7-SAAP algorithm are up to 10 times higher than for RFTA2, and 3 times higher than for RFTA2GE. BFS-7 has a similar lifetime but with higher communication overhead. BFS-10 has a lifetime twice longer than BFS-7 but with twice higher communication costs. Compared to RFTA algorithms, BFS-7 and BFS-10 has a lifetime 6 times smaller and almost 30 times higher communications costs. It can be concluded that RFTA2 and RFTA2GE outperform the k-SAAP and BFS algorithms using the combination of the face routing combined with auctions.

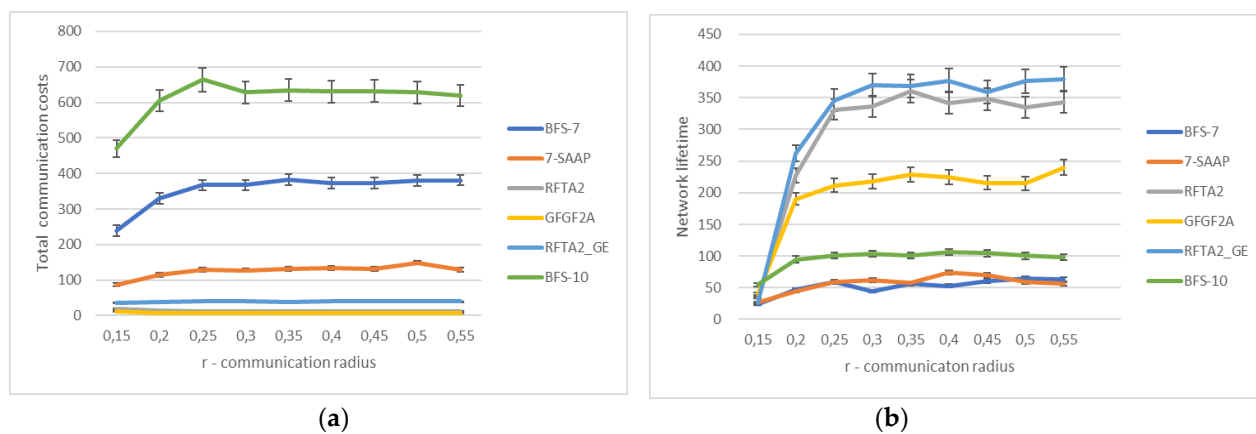

**Figure 10**: Comparison of algorithms in terms of network lifetime and commnucation costs for random network topology for (**a**) network lifetime (**b**) total communication costs

On the Figures 11 and 12 we show the obtained results for proposed algorithms in details for two different topologies and comparisons among proposed algorithms.

Figure 11 shows the average network lifetime obtained for the new RFTA2 algorithm for different radii from SR to 4SR and communication radius $r$ was varied from 0.15 to 0.55. It is shown that the longest average lifetime is for the 2SR, which also confirms results from the Section 4.5. As the communication radius $r$ increases, the lifetime becomes constant starting with r > 0.20. When the communication radius is smaller, the network is mostly disconnected and there are fewer possible routing paths, hence shorter lifetime. For the larger r, the communication radius has no more influence on network lifetime. The longest achieved lifetime is around 350 rounds. The communication costs for different search radius from SR to 4SR and communication radius $r$ varied from 0.15 to 0.55 are given in Figure 12. Communication costs are similar for both topologies.

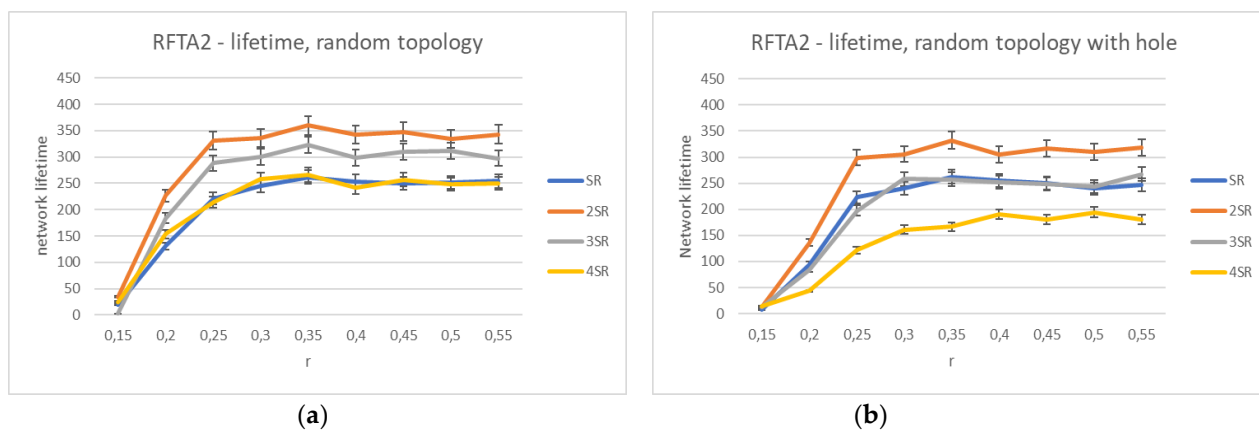

**Figure 11** Network lifetime for the RFTA2 algorithm for (**a**) random topology (**b**) random topology with hole

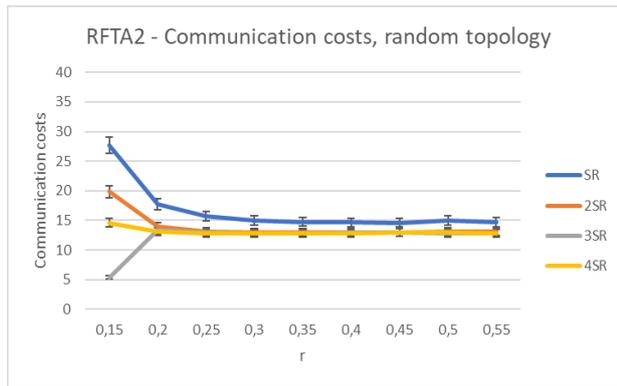
(a)
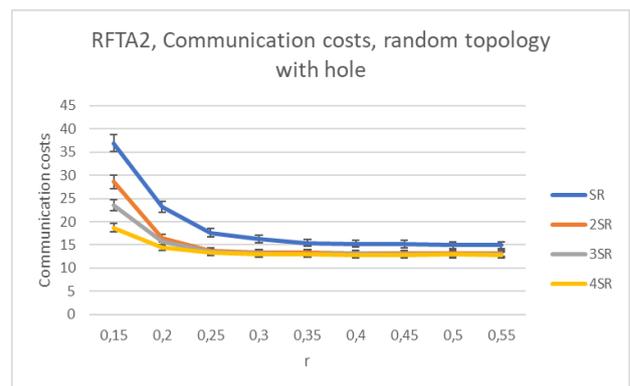
(b)

**Figure 12** Communication costs for the RFTA2 algorithm for (**a**) random topology (**b**) random topology with hole

For the greedy extension of RFTA2 algorithm, the same set of simulations is executed, and the same performance indicators are measured. The results for the network lifetime are given in Figure 13, showing that it can take up to almost 400 rounds (a little bit lower for the topology with hole). Communication costs are given in Figure 14, showing that it is around 40 messages and similar for both topologies.

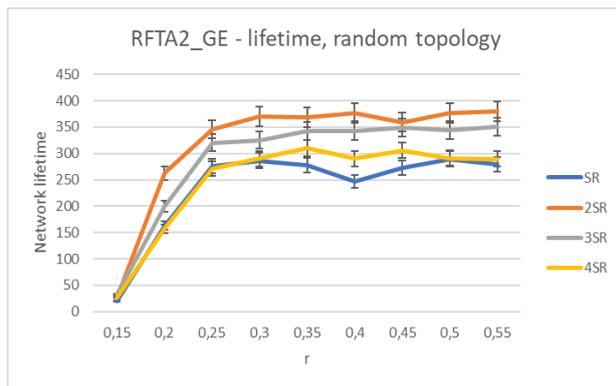
(a)
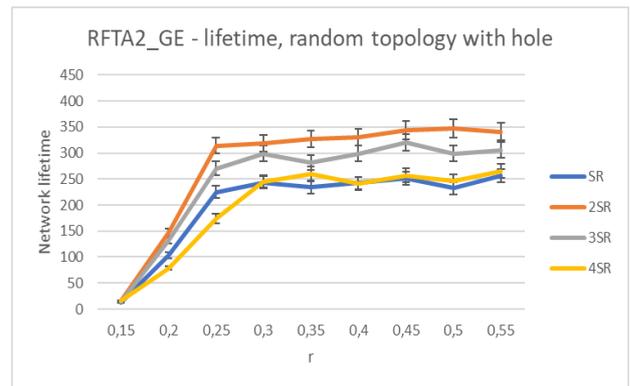
(b)

**Figure 13** Network lifetime for the RFTA2GE algorithm for (**a**) random topology (**b**) random topology with hole

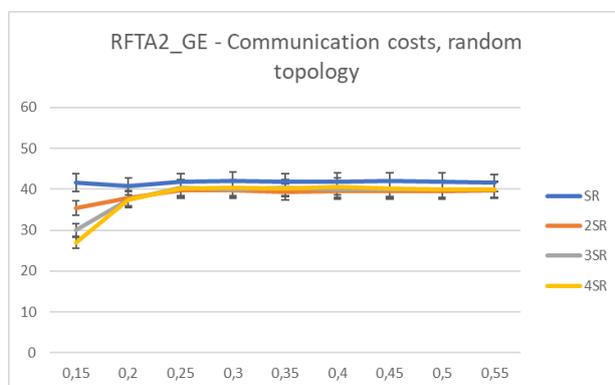
(a)
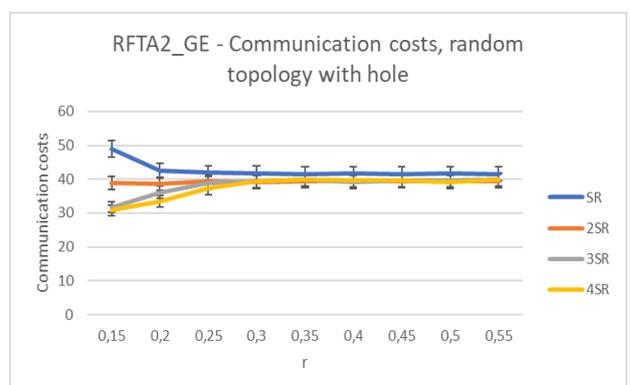
(b)

**Figure 14** Communication costs for the RFTA2GE algorithm for (**a**) random topology (**b**) random topology with hole

The simulations showed that the best results are obtained for the search radius of 2SR (as concluded in [7] also) as explained in the Section 4.5. Accordingly, communication costs and network lifetime are compared for the RFTA2, GFGF2A, and RFTA2GE for fixed search radius 2SR. From Figure 15 it can be seen that, as expected, the highest communication costs are for the RFTA2GE (i.e. greedy extension of RFTA2) algorithm for both topologies. It is around four times higher than the other two algorithms. On the other hand, it can be seen from Figure 16 that the network lifetime of the new algorithms RFTA2 and RFTA2GE is almost doubled compared to the GFGF2A algorithm for random topology scenario, and it is almost four times longer for random topology with the hole. If we compare RFTA2 and RFTA2GE, communication costs for greedy extension are much higher for a very little benefit gained in network lifetime.

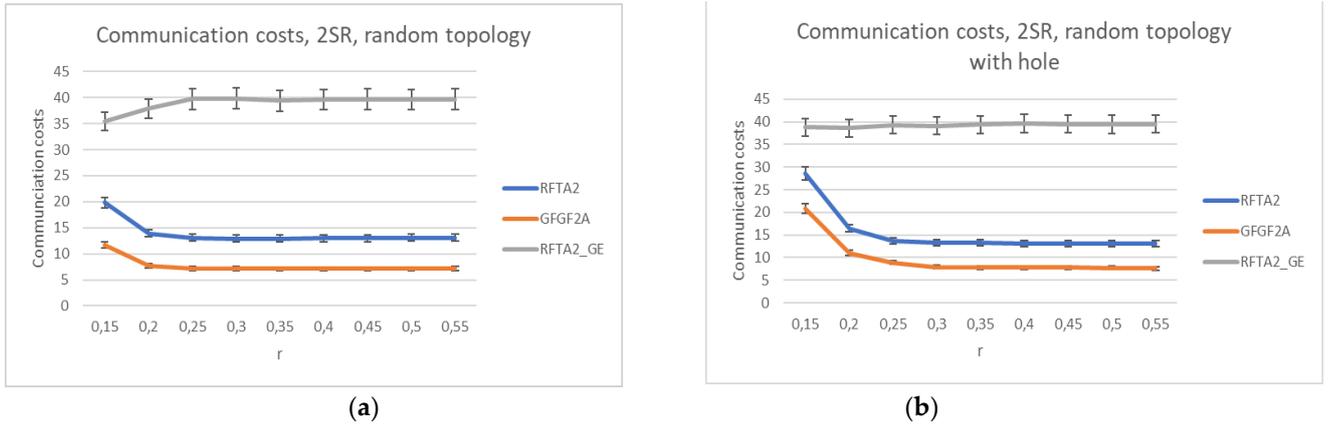

**Figure 15** Communication costs comparison for RFTA2, GFGF2A and RFTA2GE for (**a**) random topology (**b**) random topology with hole

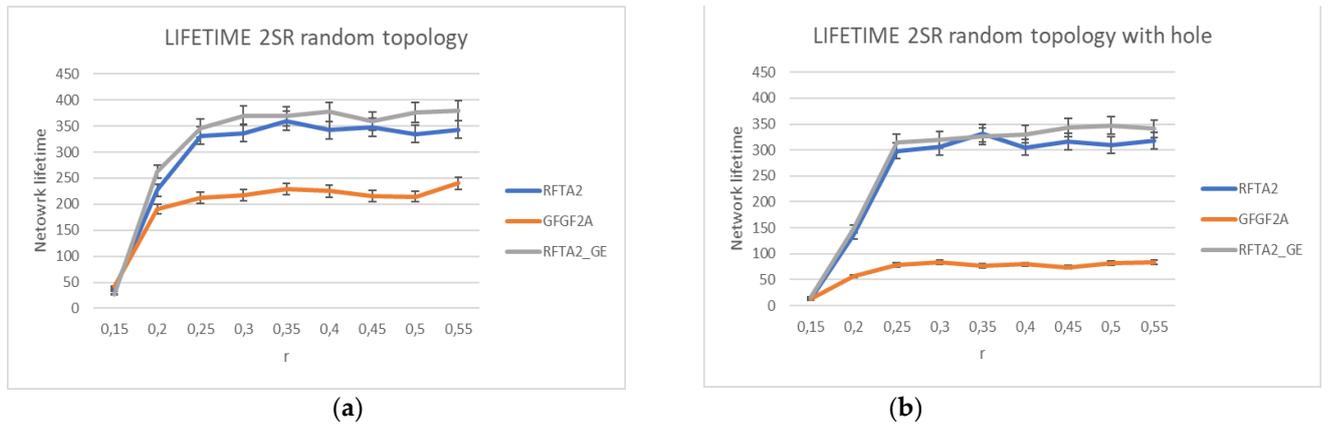

**Figure 16** Network lifetime comparison for RFTA2, GFGF2A and RFTA2GE for (**a**) random topology (**b**) random topology with hole

Furthermore, what is shown in the Figure 16 is that the lifetime for GFGF2A for the random topology with hole is four times shorter than for the random topology. The explanation is as follows. The network changes more in the case of topology with hole since the hole is occupying a significant part of the area while a task can be found anywhere within a unit square. Thus, a robot doing the task could be moving a lot and would therefore lose lots of energy. On the other hand, in the RFTA2 and RFTA2GE (having auctions), the robots with more energy are chosen and hence the hole in topology does not influence much the network lifetime.

*5.1 Robot energy statistics and balancing*

Another measured parameter is the energy each robot has been left with after the round in which the network is considered not alive (i.e. network lifetime is achieved). The Figure 17 shows two examples, one for RFTA2 and one for RFTA2GE algorithm. In both cases, the routing started on the same network. RFTA2GE features a longer network lifetime (as expected), and the energies are also better balanced.

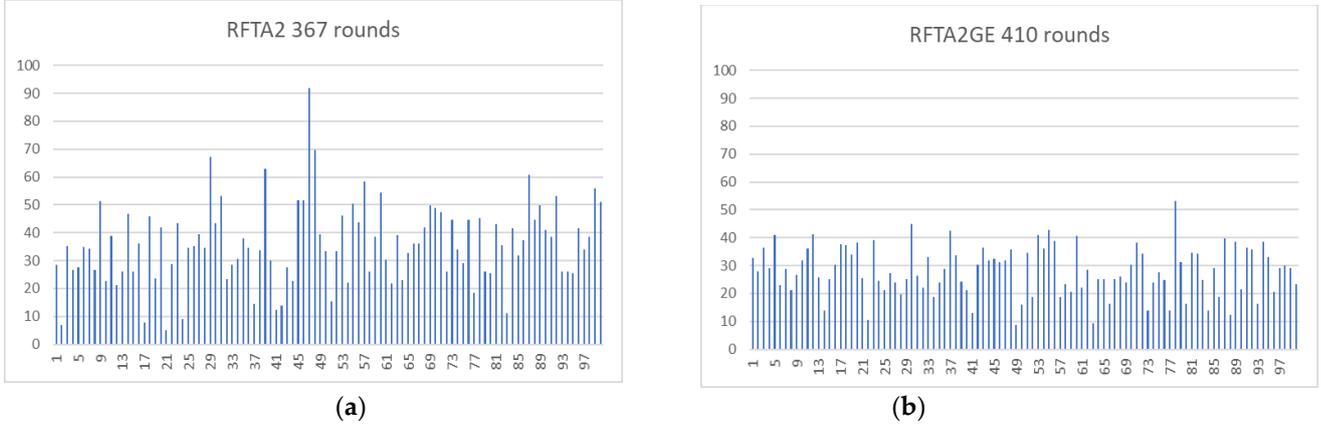

**Figure 17** Robot energies after the lifetime at the random topology (R=0.2, r=0.25)

Figure 17 shows that there are still many robots with the significant remaining energy. As the robot energy dries out, the probability that the network will be disconnected rises. However, that scenario is out of the scope of this work. There is room for further optimizations, some of which we identify in the conclusion section.

The following variables are measured with standard deviation to better explore the energy statistics and balancing among robots for all three algorithms. Average messages per robot variable is a measure of how many messages each robot has sent during the network lifetime. The difference related to the communication costs variable measuring the number of messages sent during the *single* routing task is to be noted. Other averaged variables presented are the network lifetime, minimum of remaining robot energy, remaining robot energy, number of robot reactions, and traveled distance per robot. In Tables 2 and 3 are shown obtained statistics for random topology and topology with hole.

**Table 2** Robot energies statistics after the final rounds for the random topology

|  | GFGF2A | RFTA2 | RFTA2GE |
|---|---|---|---|
| Average messages per robot (AMPR) [#msg] | 14.79±5.51 | 43.32±12.23 | 68.64±15.39 |
| Average network lifetime (ANL) [#rounds] | 225.11±88.51 | 323.14±122.69 | 376.35±96.94 |
| Average MIN of remaining robot energy [%] | 7.39 | 11.33 | 5.95 |
| Average remaining robot energy (ARRE) [%] | 79.52±6.98 | 44.81±20.17 | 35.04±13.98 |
| Average number of robot reactions (ANRR) | 2.13±0.81 | 3.07±1.14 | 3.57±0.84 |
| Average traveled distance per robot (ATDPR) [m] | 1.06±0.36 | 2.85±1.04 | 3.35±0.72 |

**Table 3** Robot energy statistics after the final rounds for the random topology with hole

|  | GFGF2A | RFTA2 | RFTA2GE |
|---|---|---|---|
| Average messages per robot (AMPR) [#msg] | 6.45±3.63 | 13.62±6.29 | 57.62±18.13 |
| Average network lifetime (ANL) [#rounds] | 85.60±57.25 | 260.16±123.88 | 319.87±105.66 |
| Average MIN of remaining robot energy [%] | 11.49 | 12.00 | 8.03 |
| Average remaining robot energy (ARRE) [%] | 90.56±5.48 | 53.44±20.76 | 41.93±16.88 |
| Average number of robot reactions (ANRR) | 0.82±0.56 | 2.47±1.15 | 3.05±0.95 |

| Average traveled distance per robot (ATDPR) [m] | 0.49±0.28 | 2.39±1.07 | 2.99±0.87 |
|---|---|---|---|

It can be concluded that the RFTA1 algorithm is not better compared to the GFGF2 algorithm in terms of finding the closest robot. Moreover, it has around 25% higher communication costs.

RFTA2 is better compared to GFGF2A in terms of longer network lifetime since it features 1.6 times longer network lifetime with 1.8 times higher communication costs for the random topology and 3.75 times longer lifetime with 1.8 times higher communication costs for the random topology with hole. The greedy extension of RFTA2 has little benefit on the network lifetime compared with RFTA2 (around 7%) with around 2.5 times higher communication costs.

For the random topologies, RFTA2GE features the best robot utilization and energy balancing of all algorithms. It features more than twice better robot utilization compared to GFGF2A and 0.8 times better compared to RFTA2 in terms of average remaining robot energy. Besides, better robot utilization can be seen from the values of the average number of robot reactions and traveled distance per robot. RFTA2GE is 1.7 times better than GFGF2A, and 1.1 times better than RFTA2 in terms of robot reactions. It is more than three times better when compared to GFGF2A and 1.2 times better compared to RFTA2 in terms of traveled distance per robot.

For the random topologies with hole, the results show that both RFTA2 and RFTA2GE outperform GFGF2A in terms of all measured variables. RFTA2GE is almost 50% better than GFGF2A and 30% better than RFTA2 in terms of average remaining robot energy. When it comes to robot reactions, the RFTA2GE is more than nine times better than GFGF2A, and six times better than RFTA2. The same trend continues for traveled distances per robot whereby it shows six times better performance than GFGF2A and 1.25 times than RFTA2.

## 6. Conclusions

In this paper we proposed four new algorithms for the robot task allocation problem in wireless sensor and robot networks in scenarios with random and random with hole topologies – the RFTA1, RFTA2, GFGF2A, and RFTA2GE. We compared between our algorithms with two similar algorithms from the literature (k-SAAP and BFS) used for task assignment. The RFTA2 and RFTA2GE feature a lifetime up to 7 times longer with significant communication overhead reduction compared to k-SAAP and BFS. The RFTA1 algorithm does not show any benefit compared to the GFGF2 algorithm. The algorithms RFTA2 and RFTA2GE were compared to GFGF2A and simulation results showed that RFTA2 is better than GFGF2A in terms of network lifetime with a little communication overhead (75% longer network lifetime for additional around ten messages higher overhead). That is particularly beneficial for the random networks with hole with the same communication overhead (4.5 times longer network lifetime). RFTA2GE (greedy extension of RFTA2) features good energy balancing, has the best network lifetime (up to 3.75 times compared to GFGF2A and up to 1.2 times compared to RFTA2), and shows solid robot utilization due to lower remaining robot energies (more than 50% lower compared to GFGF2A and around 30% lower compared to RFTA2GE) for the price of additional communication overhead. The last results are also confirmed measuring the average number of robot reactions and the average traveled distance per robot. The fact that the communication costs are usually by order of magnitude lower than robot movement costs, is a highly beneficial feature of RFTA2GE.

For RFTA2 and RFTA2GE, using the geometric probability approach given by Lemma 1 and Lemma 2, we explained why the best results for network lifetime are obtained for SR=2SR=0,2. In Lemma 3 and 4 we showed that routing complexity is upper bounded by $O(n*c)$ and communication complexity is upped bounded by $O(n^2+n)$, respectively.

One of the limits of this work is that there is still a considerable amount of robot residual energy after the network lifetime. It seems that there is still room for further improvement. Besides, the bidding metrics used by Eq. (1) could be variated with some other metrics for further comparisons. The following metrics might be used in that direction, $m*d/E_R$ or $m*d(E_R-m*d)$, where m is constant, d is distance and $E_R$ is the robot's remaining energy.

For future work, the energy loss calculation and speed of the robot could be varied to examine what impact it would have on the results and whether more benefits could be found. Besides, it would be interesting to explore how to balance the energy of robots better while maintaining the network connected. It is planned for future research to examine the scenarios with disconnected networks and to find out the ways for robot relocation to maintain the connectivity, prolonging network lifetime. The similar idea is given in [31].